\documentclass[twocolumn,showpacs,prl]{revtex4}

\usepackage{graphicx}

\usepackage{amssymb}

\usepackage[center]{subfigure}
\begin{document}

\def\Re{\textrm{Re}}

\title{The friction factor of two-dimensional rough-boundary turbulent soap film flows}
\author{Nicholas Guttenberg and Nigel Goldenfeld}
\affiliation{Department of Physics, University of Illinois at
Urbana-Champaign, 1110 West Green Street, Urbana, Illinois, 61801-3080.}

\begin{abstract}
We use momentum transfer arguments to predict the friction factor $f$
in two-dimensional turbulent soap-film flows with rough boundaries (an analogue
of three-dimensional pipe flow) as a function of Reynolds number $\Re$ and
roughness $r$, considering separately the inverse energy cascade and
the forward enstrophy cascade.  At intermediate $\Re$, we predict a
Blasius-like friction factor scaling of $f\propto\textrm{Re}^{-1/2}$ in
flows dominated by the enstrophy cascade, distinct from the energy
cascade scaling of $\Re^{-1/4}$. For large $\Re$, $f \sim r$ in
the enstrophy-dominated case. We use conformal map techniques to
perform direct numerical simulations that are in satisfactory agreement
with theory, and exhibit data collapse scaling of roughness-induced
criticality, previously shown to arise in the 3D pipe data of
Nikuradse.

\end{abstract}
\maketitle

Turbulent flows are marked by rich structure over a range of
scales---they host fluctuations, vortices, tangles, and other coherent
structures that continue to defy a detailed, analytical
understanding\cite{CHOR97,SREE99}.  When parameterized in terms of the
typical flow speed $U$, characteristic length scale $L$ and kinematic
viscosity of the fluid $\nu$, three-dimensional turbulence exhibits
universal phenomena as the Reynolds number $\Re\equiv UL/\nu
\rightarrow \infty$.  Most famously, in a theory referred to as
K41\cite{KOLM41,OBUK41}, the dependence of the fluctuation energy
spectrum $E(k)$ on wavenumber and mean energy transfer rate
$\bar\epsilon$ occurs in a way that is independent of $\nu$: $E(k)=
\bar\epsilon^{2/3} k^{-5/3}$ for values of wavenumber in the so-called
inertial range, intermediate between the scales of forcing and the
scales where molecular viscosity becomes significant.  In this inertial
range, turbulent eddies break up into smaller eddies through a
mechanism which is to a first approximation Hamiltonian, and results in
a cascade of energy to smaller length scales\cite{richardson1922wpn}.

During the 1930's, Nikuradse undertook a systematic series of
measurements of the pressure drop across a turbulent pipe flow as a
function of $\Re$\cite{NIKU32} and also as a function of $r/R$, the
scale of the roughness of the pipe walls $r$, normalized by the pipe
radius $R$\cite{NIKU33}. The former measurements provided strong
support for Prandtl's boundary layer concept\cite{schlichting2000blt},
and have been replicated and surpassed only
recently\cite{mckeon2004ffs}, while the latter measurements, despite
recent efforts\cite{shockling2006ret,ALLE07}, remain to this day the
most complete data set of its kind, spanning three orders of magnitude
in Reynolds number and a decade in the dimensionless roughness $r$.
These data reveal that the frictional drag experienced by a turbulent
fluid in a pipe with rough walls is a non-monotonic and complicated
function of Reynolds number and roughness, which despite intense
interest and practical importance (see, e.g. Ref.
\cite{schlichting2000blt}), has only begun to be
understood\cite{GIOI06,GOLD06} through two related developments.

First, Gioia and Chakraborty\cite{GIOI06} estimated the
momentum-transfer between the walls of the pipe and the flow,
explicitly taking into account the presence of roughness. Their
resultant formula for the dimensionless friction factor (defined
precisely below) is expressed in terms of the turbulent kinetic energy
spectrum $E(k)$, and thus makes a direct connection between a
macroscopic flow property and the velocity field correlations. Second,
Goldenfeld\cite{GOLD06} pointed out that the power law behavior of
Nikuradse's friction factor data in the regimes $\Re\rightarrow\infty$
and $r/R \rightarrow 0$ was analogous to critical phenomena, where the
inverse Reynolds number and roughness play similar roles to, for
example, the coupling constant and external magnetic field in an Ising
model. Consequently, the dependence of Nikuradse's data on $\Re$ and
$r$ can be collapsed onto a universal function with sufficient
precision for intermittency corrections to be extracted\cite{MEHR08}.
These results show that the friction factor reflects the nature of the
turbulent state through its dependence on the energy spectrum, and that
the turbulent state is itself a manifestation of a non-equilibrium
critical point at $\Re=\infty$ and $r/R \rightarrow 0$.

The purpose of this Letter is to test the claims of Refs. \cite{GIOI06}
and \cite{GOLD06} in a context where detailed calculations are in
principle possible: the case of two-dimensional soap-film
turbulence\cite{kraichnan1980tdt, KELL02}. Here, a soap film is
supported between two vertical wires, and the draining flow provides a
versatile laboratory for exploring two-dimensional
turbulence\cite{KELL02}. It is well-understood that the nature of
turbulence in 2D is different from 3D: there is no vortex stretching,
for example. Nevertheless, turbulent phenomena exist, and possess the
novelty that there are two cascades: an energy inverse cascade that
runs from small to large scales\cite{batchelor1982tht, KRAI67}, and a
forward cascade\cite{KRAI67} in the enstrophy $\Omega \equiv
|\bf{\nabla} \times \bf{v}|^2$, where $\bf{v}$ is the fluid velocity
field.  This enstrophy cascade yields an energy spectrum
$E(k)=\beta^{2/3} k^{-3}$, where $\beta$ is the rate of transfer of
enstrophy.

Prior work, dating back to Prandtl and others (for a review see
Ref.~\cite{schlichting2000blt}) is not able to make a prediction about
the friction factor in these cases, because it has no specific
representation of the nature of the turbulent state, and in particular
is disconnected from the energy spectrum.  On the other hand, the
momentum-transfer theory of Gioia and Chakraborty\cite{GIOI06} can
reflect the character of 2D turbulent states, as expressed by the
energy spectrum, through the dependence of the friction factor on $\Re$
and $r$. We show below that the momentum-transfer theory predicts a
significant dependence of the friction factor on the nature of the
turbulent cascade, one that we observe in direct numerical simulations
reported here, and which obeys the scaling predicted by
roughness-induced criticality. Thus our direct numerical calculations
agree well with the momentum-transfer and roughness-induced criticality
picture, and strongly suggest that the standard picture of turbulent
boundary layers is incomplete.

\medskip
\noindent {\it Calculation of the friction factor scaling laws in
2D:-\/} In the momentum-transfer theory of Gioia and Chakraborty, the
friction factor is shown to be proportional to the root-mean-square
velocity fluctuation $u_s$ at a scale $s$ determined by the larger of
the roughness $r$ or the Kolmogorov scale $\eta_K$.  Since $E(k)dk$
represents the turbulent kinetic energy in the wavenumber band between
$k$ and $k+dk$, it follows that:
\begin{equation}
u_s=\left[\int_{1/s}^{\infty} E(k)dk\right]^{1/2}
\end{equation}
Anisotropy near the wall has only a small effect\cite{CASC05} on the
low-order structure function used in our calculation. For simplicity,
using the K41 form for $E(k)$, we obtain $f\propto \bar\epsilon^{1/3}
s^{1/3}$.  With the limiting forms for $s$ at large and small Reynolds
number, we obtain the predictions of the empirically-observed Blasius
regime\cite{BLAS13}, in which the friction scales as  $\Re^{-1/4}$ for
small but turbulent Reynolds numbers, and the Strickler
regime\cite{STRI23} at large enough $\Re$, where the friction factor
is independent of $\Re$ and only depends on the roughness
through the relation $f\propto (r/R)^{1/3}$.

In two-dimensional turbulent systems, both the energy cascade or the
enstrophy cascade may be observed, or they may occur
individually\cite{RUTG98} depending on the manner of energy injection
and the scale at which it occurs. The two dimensional inverse cascade
friction factor is the same as the case of three dimensional flows,
with a Blasius scaling of $f \propto \Re^{-1/4}$ and a Strickler
scaling of $f\propto (r/R)^{1/3}$. The energy spectrum due to the
enstrophy cascade leads to a new prediction for the friction factor: a
scaling of $f \propto \Re^{-1/2}$ in the Blasius regime and $f \propto
(r/R)$ in the Strickler regime.  These are our central predictions,
which we seek to verify by numerical simulation in the next section.
In general, the friction factor corresponding to any conserved quantity
(such as helicity) with units $\textrm{[L]}^a \textrm{[T]}^b$ is $f
\propto \Re^{-(1-\phi)/(2-\phi)}$ (Blasius regime) and $f \propto
(r/R)^{1-\phi}$ (Strickler regime), where $\phi \equiv a/(1-b)$.

\begin{figure}[t]
\includegraphics[width=\columnwidth,angle=0]{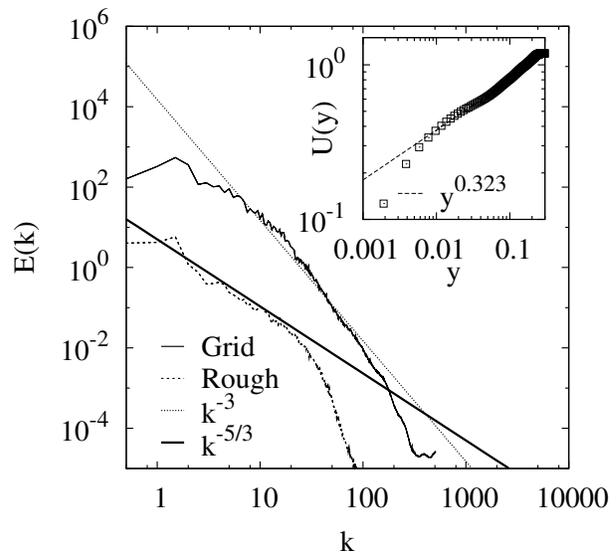}
\caption{Energy spectra for grid- and roughness-generated turbulence.
Grid-generated turbulence exhibits the $k^{-3}$ enstrophy cascade,
whereas roughness-generated turbulence exhibits the $k^{-5/3}$ inverse
cascade scaling. Inset: simulated wall velocity profile of
grid-generated turbulence in a smooth pipe at $\textrm{Re}=60000$. The
profile is consistent with a power law with exponent $0.323 \pm 0.005$.
We predict an exponent of $1/3$ for enstrophy cascade turbulence.}
\label{FigSpectra}
\end{figure}

\medskip
\noindent
{\it Simulations of 2D turbulent rough-pipe
flows:-\/} To test the momentum transfer theory's prediction of the
friction factor in 2D, we have performed simulations for a range of
Reynolds numbers and single-wavelength roughness, both with
grid-generated turbulence and turbulence generated by wall roughness.
The roughness of the wall breaks translational invariance and means
that one cannot simply solve the Navier-Stokes equations using spectral
methods.  We have overcome this difficulty by using a
judiciously-chosen conformal map technique, allowing us to use a
spectral method to satisfy incompressibility.  The SMART
algorithm\cite{GASK88} is used to calculate the advection of the
velocity field. The friction factor is measured by computing the
pressure drop necessary to maintain the average flow velocity over the
periodic domain.

To simulate a rough-walled pipe, we apply a conformal map of the form
$w=z+r \exp(ikz)$, where the aspect ratio is held constant ($rk=3/4$)
and the wavenumber $k$ may be varied to produce roughness of different
scales.  Note that $r$ plays the role of roughness in Nikuradse's
experiments, but our aspect ratio is $3/4$ and not unity as in his
experiments. This conformal map results in the addition of two body
force terms to the Navier-Stokes equation in the transformed
(rectangular) domain, in addition to an overall weighting factor
deriving from the changed volume of each cell:
\begin{equation}
|g^\prime|^2 \frac{\partial {\bf \bar{V}}}{\partial t} + ({\bf \bar{V}}
\cdot {\bf \nabla}){\bf \bar{V}} = \nu \nabla^2 {\bf \bar{V}} + \frac{
|{\bf \bar{V}}|^2 }{|g^\prime|^2} {\bf A} + \frac {2 \nu }
{|g^\prime|^2} {\bf A}^{\perp} ({\bf \nabla} \times {\bf \bar{V}})
\end{equation}

\noindent
Here $|g^\prime|^2 = x_u^2+x_v^2 = y_u^2+y_v^2 = x_u y_v - x_v y_u$,
${\bf \bar{V}}$ is the velocity in the transformed coordinates, and the
vector ${\bf A}$ is defined as:
\begin{equation}
{\bf A} \equiv \left[ \begin{array}{c} x_u y_{uv} + x_v x_{uv} \\ x_u x_{uv} - x_v y_{uv}\end{array} \right]
\end{equation}

We use a simulation domain of $2048 \times 512$ to simulate a section
of pipe of diameter $1$ and length $4$. After initializing the velocity
field we allow the system to evolve for a sufficient number of pipe
transits so that the system is fully turbulent (one pipe transit
corresponds to four units of time as the mean flow velocity is set to
$1$ in the simulation units). The smaller the roughness, the more
transits are needed. This results in roughness-generated turbulence, in
which case the observed energy spectrum is dominated by the inverse
cascade, as shown in Fig.~(\ref{FigSpectra}).

In order to attain an enstrophy-dominated flow, we used a technique
suggested by the observations reported by Rutgers\cite{RUTG98}.  We
simulated grid-generated turbulence, by placing a series of cylinders
at the mouth of the pipe; in each cylinder we set the velocity field to
zero every timestep. After one pipe transit the velocity field is fully
developed. We then remove the grid and allow the turbulence to decay
for a transit before we begin to measure the friction factor and other
flow properties. We have observed energy spectra dominated by the
enstrophy cascade in this system, as shown in Fig.~(\ref{FigSpectra}).

\begin{figure}[t]
\includegraphics[width=\columnwidth,angle=0]{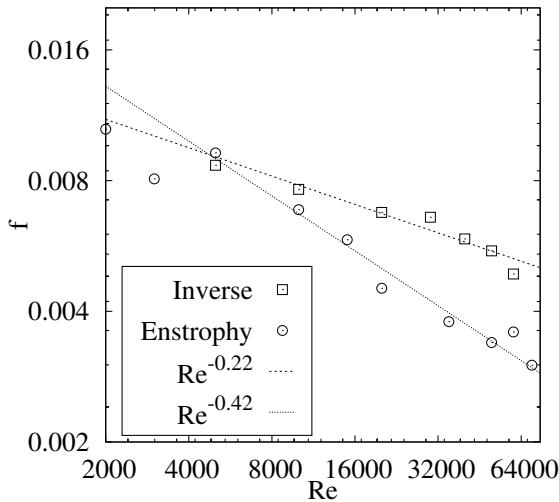}
\caption{Scaling of the friction factor with respect to $\Re$
for inverse cascade and enstrophy cascade dominated flows in 2D.  The
roughness is $r/R= 0.067$, and the data have been averaged over a time
of 5 pipe transits.} \label{FigFFactor}
\end{figure}

Our simulation results at small values of the dimensionless roughness
($r/R= 0.067$) are plotted in Fig.~(\ref{FigFFactor}).  These results
were obtained by averaging over 5 full pipe transits, yielding
reproducible values for the friction factor, with controlled error
bars, as shown. For this flow we observe an approximate power-law
scaling of the friction factor with Reynolds number, with an exponent
$-0.22\pm0.03$ together with an energy spectrum dominated by the
inverse-cascade. In the case of grid-generated decaying turbulence,
corresponding to an enstrophy-cascade dominated spectrum, we observe an
exponent of $-0.42\pm0.05$. These results are within satisfactory
agreement with the scalings of $-1/4$ and $-1/2$ respectively,
predicted for the 2D Blasius regime on the basis of a momentum transfer
argument.

\begin{figure}[t]
\includegraphics[width=\columnwidth,angle=0]{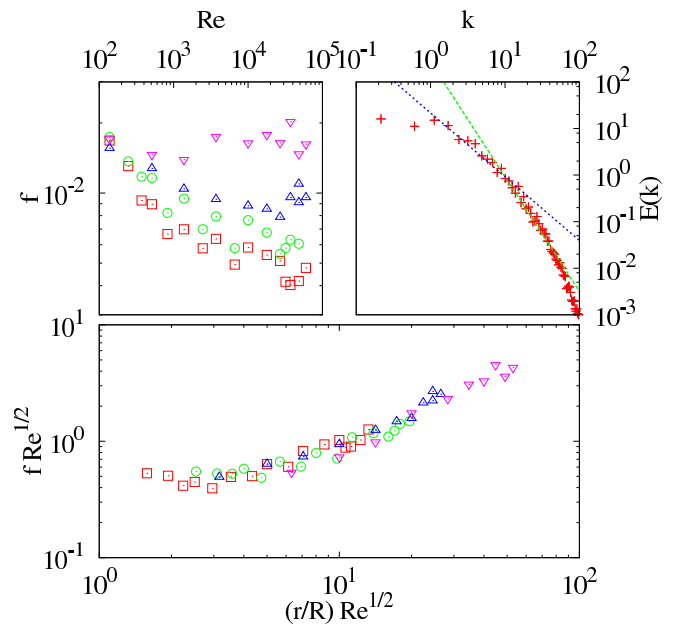}
\caption{(Color online). The bottom inset shows the enstrophy cascade
data collapse of the friction factor curves for nondimensional
roughness $0.05$ ($\circ$), $0.08$ ($\square$), $0.1$ ($\triangle$),
and $0.2$ ($\triangledown$) over a range of Reynolds numbers from
$1000$ to $80000$. The top left insent shows the unscaled friction
factor data. The top right inset shows the energy spectrum at
$r/R=0.08$ and $\Re=80000$. The straight lines correspond to
$k^{-5/3}$ and $k^{-3}$.} \label{FigDataCollapse}
\end{figure}

We cannot reach sufficiently high Reynolds numbers to observe a pure
Strickler regime, but we can verify the Strickler scaling exponent with
data collapse. In three dimensions, or in a system dominated by the
inverse cascade, we expect data collapse when plotting $f
\Re^{1/4}$ against $(r/R) \Re^{3/4}$\cite{GOLD06}. For
the enstrophy cascade, these variables should be $f \textrm{Re}^{1/2}$
and $(r/R) \Re^{1/2}$ respectively. We have observed previously
that in the presence of roughness, the spectrum is dominated by the
inverse cascade. However, we have found that by adding a small amount
of random forcing to the velocity field, the enstrophy cascade may be
observed even in a rough pipe, though it may be coexistant with an
inverse cascade. Using this method we can obtain the roughness
dependence of the friction factor in an enstrophy cascade dominated
flow. The collapse of the friction factor curves using the enstrophy
cascade variables is shown in Fig.~(\ref{FigDataCollapse}). The
collapse is quite good, despite an apparent shallowness to the Blasius
regime in the raw data. This shallowness is likely caused by the
presence of a small amount of roughness, modifying the expected
$\Re^{-1/2}$ scaling at larger Reynolds numbers. We have
neglected intermittency, which is negligible in 2D\cite{paret1998itd}.

\medskip
\noindent {\it Relationship of the friction factor to the velocity
profile:-\/} Following Prandtl\cite{PRAN21}, we have calculated the
mean velocity profile $u(y)$ as a function of distance from a wall $y$,
and for the enstrophy cascade this yields $u(y) \sim y^{\alpha}$ with
$\alpha=1/3$, corresponding to the Blasius regime. For a general
conserved quantity, $\alpha = (1-\phi)/(3-\phi)$. This relation depends
on the zero roughness limit. In \cite{KOTE03}, it has been shown that
roughness modifies the velocity profile so as to increase the apparent
scaling exponent. Other work\cite{PATE98,AFZA06} also considers the
influence of rough walls on the velocity profile and near-wall scaling.

In our simulations of smooth-pipe enstrophy cascade turbulence, we have
measured the velocity profile and found the power-law scaling exponent
$\alpha=0.323\pm0.005$ between $0.01 R$ and $0.1 R$, as shown in the
inset of Fig.~(\ref{FigSpectra}), close to the predicted $\alpha=1/3$.
In the case of our rough-pipe simulations, the velocity profile yielded
an exponent of $0.333\pm0.002$, significantly steeper than the
predicted $\alpha=1/7$  that applies in the smooth, inverse cascade
case. Our interpretation is that this is due to spectral contamination
from an enstrophy cascade, as in the case of the simulations with
random forcing that we presented. The momentum transfer theory integral
has an upper limit that is comparable with the Kolmogorov lengthscale
at low roughness, and so in that case the small-$k$ part of the energy
spectrum controls the friction factor scaling. Because of this, we
would expect to see a velocity profile consistent with the enstrophy
cascade until the roughness or Reynolds number were high enough to
place the crossover between the inverse cascade and contaminant
enstrophy cascade below the scale of the roughness.

Our results for the power-law Blasius regime in 2D enstrophy-dominated
turbulence show convincingly that this regime is more than an empirical
fit, and has a dynamical significance. Our direct numerical simulations
support the fundamental connection between spectral structure and
friction factor scaling, which is manifested in the observed
roughness-induced criticality.

We are grateful for valuable discussions with Gustavo Gioia, Pinaki
Chakraborty, Carlo Cesar Zuniga Zamalloa, Patricio Jeraldo, Tuan Tran,
John Kolinski, Hamid Kellay and Walter Goldburg.  Nicholas Guttenberg
was partially supported by a University of Illinois Distinguished
Fellowship. This material is based upon work supported by the National
Science Foundation under Grant No. NSF DMR 06-04435.

\bibliographystyle{apsrev}
\bibliography{turbulence}
\end{document}